\documentclass[onecolumn,superscriptaddress,secnumarabic,amssymb,amsmath,nobibnotes,aps,prd,nofootinbib,altaffilletter]{revtex4}
\usepackage{amsmath,amssymb}
\usepackage{graphicx}
\usepackage{dcolumn}
\usepackage{bm,color}
\usepackage{hyperref}
\usepackage{accents}
\usepackage{amssymb,float}
\usepackage{amsmath}
\usepackage{multirow}
\usepackage{siunitx}
\usepackage{tabularx}
\usepackage{booktabs}
\usepackage{url}
\usepackage{orcidlink}

\usepackage{capt-of} 
\usepackage{placeins} 
\usepackage{psfrag}
\usepackage{times}
\usepackage[varg]{txfonts}
\usepackage{xcolor}
\usepackage{xspace}

\begin{document}


\title{Chameleon perfect scalar field as a geometric correction in $f(R)$ gravity}

\author{Laura L. Parrilla}
\email{laura.parrilla@correo.nucleares.unam.mx}
\affiliation{Instituto de Ciencias Nucleares, Universidad Nacional Aut\'{o}noma de M\'{e}xico, 
Circuito Exterior C.U., A.P. 70-543, M\'exico D.F. 04510, M\'{e}xico.}

\author{Celia Escamilla-Rivera\orcidlink{0000-0002-8929-250X}}
\email{celia.escamilla@nucleares.unam.mx}
\affiliation{Instituto de Ciencias Nucleares, Universidad Nacional Aut\'{o}noma de M\'{e}xico, 
Circuito Exterior C.U., A.P. 70-543, M\'exico D.F. 04510, M\'{e}xico.}

\begin{abstract}
In this work, we derive the analytical form for a $f(R)$ model that describes a perfect scalar field $\phi$ by assuming the existence of a chameleon mechanism. 
Based on four statements, at the background and perturbative level, it is possible to relate the extra terms from this theory as a geometrical perfect fluid term, 
whose has been expressed as possible candidates to explain the nature of the dark sector, and possibly, in the case of a perfect scalar chameleon during inflation, satisfy the big bang nucleosynthesis (BBN) constraints until late times.
\end{abstract}


\maketitle
\section{Introduction} 
\label{sec:intro}

Since the discovery of the current accelerated cosmic expansion of the Universe using observational data from astrophysical objects like Supernovae type Ia independently by the high-redshift Supernovae Search Team \cite{SupernovaSearchTeam:1998fmf} and the Supernovae Cosmology Project Team \cite{SupernovaCosmologyProject:1998vns}, has been attributed the source of this phenomena to the so-called dark energy. However, even with the many efforts to pursue the understanding of the nature of such a dark component, this has not been observationally identified yet. One of the main characteristics of dark energy is attributed to a negative pressure, which leads to an accelerated expansion phenomenon by counteracting the force of gravity. The fact that the negative pressure leads to the cosmic acceleration may look counter-intuitive\footnote{In Newtonian gravity the pressure is related to a force associated with a local potential that depends on the position in space.}. The time-dependent pressure $p(t)$ in homogeneous and isotropic spacetimes appears in general relativity (GR) and mechanisms that generate this negative pressure and cosmic acceleration are one of the main research topics in Cosmology.

The straightforward candidate proposed for dark energy is the cosmological constant $\Lambda$, whose energy density remains constant and allows Einstein field equations to preserve the conservation of energy. The $\Lambda$ term can be interpreted as a \textit{perfect} fluid by shifting to the right-hand side (\textit{r.h.s})\footnote{There are several alternative models of dark energy also on the \textit{r.h.s} of the field equations, $\Lambda$ being the simplest of all of them. }, and if dark energy modelled as $\Lambda$, for which equation-of-state $w_{\text{fld}}=-1$, is interpreted as a perfect fluid with $p_\Lambda \propto -\rho_{\Lambda }$.
From the point of view of particle physics, the Cosmological Constant can be related to a vacuum energy density, where if we sum up zero-point energies of all normal modes of some field and consider the cut-off of the momentum at Planck scale, the vacuum energy density is around $10^{121}$ times larger than the observed $\Lambda$ density, the so-called vacuum catastrophe. On the one hand, if the Cosmological Constant is truly the consequence of the present current cosmic acceleration, we need to find a mechanism to obtain the tiny value of $\Lambda$ which could be consistent with observations. 

On the other hand, if the origin of dark energy is not the cosmological constant, we need to search for some alternative (or extended) gravity models to explain the current cosmic acceleration. There are two approaches to building dark energy models other than using $\Lambda$.
The first approach is to modify the \textit{r.h.s} of Einstein field equations, ($G_{\mu\nu}=8\pi GT_{\mu\nu}$), by considering specific forms of the energy-momentum tensor $T_{\mu\nu}$ that includes the possibility of a fluid with negative pressure. The most popular models that belong to this class are the so-called quintessence \cite{Caldwell:2000wt}, k-essence \cite{Rendall:2004kh} and \textit{perfect} fluid models \cite{Faraoni:2022gry}. On this path, the methodology to follow is to consider the existence of scalar fields\footnote{In quantum field theory, several species of elementary particles corresponding to a field are produced. The fields are classified as boson or fermionic depending on the spin of the particle. In the case of scalar fields, this spin has a value of zero. At the quantum level, each field corresponds to an operator, however, bosonic fields can be considered classical in a suitable regime.} with slowly/smooth varying potentials, whereas in k-essence it is the scalar field kinetic energy that drives the acceleration. The latter class is based on a perfect fluid of a specific equation-of-state. 

The second approach to constructing dark energy models is to modify the \textit{l.h.s} of Einstein field equations. We denote the representative models that belong to this class as \textit{modified gravity} \cite{Clifton:2011jh}. While scalar field models correspond to a modification of the energy-momentum tensor, the approach in this path corresponds to the modified gravity in which the gravitational theory is modified compared to general relativity. For example, the Lagrangian density for general relativity is given by $f(R)=R-2\Lambda$, where $R$ is the Ricci scalar and the constant allows us to have the acceleration phenomena required.  A possible modification of this scenario can be described by a non-linear arbitrary function $f$ in terms of $R$, which is called $f(R)$ theories of gravity. Dark energy models based on these kinds of theories have been studied extensively including metric formalism \cite{Capozziello:2002rd,Capozziello:2003gx}, observational test \cite{Amendola:2006we,Bajardi:2022ocw,Jaime:2018ftn} and modifications to the spectra of galaxy clustering \cite{Capozziello:2017zow,DeMartino:2013zua}.

Historically, Brans-Dicke theory, an important class of scalar-tensor theories, gives rise to a constant coupling $Q$ between the scalar field $\phi$ and the matter component. In this sense, modified gravity theories can be regarded as a coupled quintessence scenario in Einstein's frame. In the absence of a scalar field potential $V(\phi)$, the Solar system tests constraint the strength of this coupling Q to be smaller than the order of $\approx 10^{-3}$. In this case, is not possible to satisfy the local gravity constraints unless we have a $V(\phi)$ with a large mass that can be capable to suppress the $Q$ coupling in the regions of high density. Furthermore, if the same field $\phi$ is responsible for the current cosmic acceleration phenomena, the potential $V(\phi)$ needs to be sufficiently flat in the regions of low density. On this line of thought, 
these requirements are possible to fulfill for large coupling models that satisfy the local gravity constraints through the chameleon mechanism \cite{Khoury:2003rn}.
The existence of a matter coupling gives rise to an extremum of the scalar field potential where the field can be stabilised. In high-density regions, such as the interiors of astrophysical objects, the field mass would be sufficiently large to avoid the propagation of the fifth force. Meanwhile, the field would have a much lighter mass in low-density environments, far away from compact objects, so it could be responsible for the present cosmic acceleration. In this work, we are going to consider a particular class of the chameleon field with an inverse power-law potential of the form $V(\phi) = M^{4+n} \phi^{-n}$, with $n\geq 1$, so local gravity constraints can be satisfied for M $\sim 10^{-2}$eV. Interestingly, these constrictions correspond to the energy scale required for the current cosmic acceleration observed. 

Recently, the correspondence between the modified gravity geometrical terms inspired by perfect fluid components has been pointed out as a natural way to understand the nature of dark energy, and by extension, the full dark sector \cite{Unnikrishnan:2010ag,Capozziello:2022sbo}.  In resume, beyond the Ricci curvature scalar $R$ or any other geometric invariant in a gravitational action, we can modelled these extra terms as perfect fluids. As an extension of these ideas, in this work, we propose to extend the \textit{conditions} that allow us to model the scalar field as perfect fluids using a chameleon scalar field. 
Perfect fluids play a crucial role in general relativity being the natural sources of Einstein field equations compatible with Bianchi identities. This characteristic allows any source of field equations to be reformulated in adequate perfect fluid form, in principle, to solve the dynamics related to the Cauchy problem. In Cosmology, perfect fluids can represent, at least in a coarse-grained image Hubble's effective flow behaviour ranging from inflation to dark energy epochs. For these reasons, the compatibility of perfect fluid solutions with modified (or extended) theories of gravity is a crucial topic to be investigated.

This paper is divided as follows: 
In Sec.\ref{sec:background} we discuss the perfect fluid form scheme for a FLRW spacetime and the three conditions to fulfill to have the perfect fluid description relationship.
In Sec.\ref{sec:perturbations} we present a fourth condition at a perturbative level to describe a perfect scalar field.
In Sec.\ref{sec:geomPF} it is described the $f(R)$ theory at the background and perturbative level.
In Sec.\ref{sec:chameleon} we introduce the chameleon scalar field in $f(R)$ theories and the analytical form of $f(R)$ to obtain a perfect scalar field. 
Finally, our discussions are given in Sec.\ref{sec:discussions}.

\section{Background conditions: perfect scalar field in FLRW spacetimes}
\label{sec:background}

We can define a FLRW spacetime by considering a zero Weyl tensor, $C_{jklm}=0$, with a time-like unit vector field in a covariant configuration as $u^ku_k=-1$ \cite{Mantica:2016iun}, as
\begin{eqnarray}
\label{eq:FLRW}
ds^2 =-dt^2 + a^2 (t) \left[ \frac{dr^2}{1-kr^2} +r^2 (d\theta^2 +\sin^2 \theta d\phi^2)\right],
\end{eqnarray}
in spherical coordinates and $k=0,-1,+1$.
Under such a description, we can compute the following covariant derivative
\begin{equation}
\label{eq:H}
\nabla_j u_{k}= H(g_{ij}+u_{j}u_{k}), \quad     \nabla_j H= -u_{j} u^{k} \nabla_{k} {H},
\end{equation}
where $H\equiv \dot{a}/a$. Here we are considering the notation $\mu,\nu =1,2,3$ and $i,j,k,l,m = 0,1,2,3$.
In a previous analysis \cite{Capozziello:2022sbo} was considered the approach given by $h_{jk}=g_{j k} + u_{j}u_{k}$, as the projection on the vector space locally orthogonal to $u_k$, therefore, we can compute the Riemann tensor from $R_{jklm}u^{m}=(\nabla_j\nabla_k-\nabla{k}\nabla_{j}) u_{l}$, and its contraction, the Ricci tensor as
\begin{equation}
\label{eq:Ricci-perfect}
R_{kl}=\frac{1}{3}\left(R-4\zeta \right)u_{k}u_{l}+\frac{1}{3}\left(R-3\zeta\right)g_{kl},
\end{equation}
where $R=R^{k}R_k$ is the curvature scalar and $\zeta= 3\ddot{a}/a$. Notice that this form of the Ricci tensor has a perfect fluid form since the terms related to the time-like vector and the shear/vorticity/acceleration-free term are separated.

To reach a perfect fluid form like in Eq.(\ref{eq:Ricci-perfect}), we need to consider the following three geometrical background tests on each of the parameters involved \cite{Capozziello:2022sbo} :
\begin{enumerate}
\item \textit{If $\phi$ is perfect, therefore a function $f(\phi)$ is perfect as long as $f$ is smooth. This characteristic applies also to products and time-derivatives of such functions, e.g. if $H$ is perfect, $H^2$ and $\dot{H}$ are perfect.}
\item \textit{Considering (1), the derivatives of a perfect scalar field $\phi$, $\nabla_{i}\nabla_{j}{\phi}$, also has a perfect fluid form. }
\item \textit{Under (1) and (2), we conclude from Eq.(\ref{eq:Ricci-perfect}) that $R$ is a perfect scalar.}
\end{enumerate}

\section{Scalar perturbation conditions: non-adiabatic pressure}
\label{sec:perturbations}

While the three conditions described in the latter section restrict us to work with perfect scalar forms, we need to be sure that such statements are also conserved at the perturbative level. In this way, we need to consider the FLRW (\ref{eq:FLRW}) as a perturbed metric:
\begin{eqnarray}
ds^2 &=& (1+2A)dt^2 -2a(t) (\partial_i B) dt dx^i 
 - a^2{t} \left[ (1-2\psi)\delta_{ij} +2(\partial_i \partial_j E)dx^i dx^j\right],
\end{eqnarray}
where A, B, $\psi$, and $E$ will be related to the Bardeen functions in the standard scalar perturbations \cite{Bardeen:1980kt}. From this point forward we are going to consider linear order perturbations, therefore, we can derive the standard conformal time coordinate potential equation as:
\begin{eqnarray}
\label{eq:bardeen}
&& \Psi'' -3\mathcal{H} (1+c^2_a) \Psi' -c^2_a \nabla^2\Psi 
+ [2\mathcal{H}' + (1+3c^2_a)\mathcal{H}^2]\Psi 
= (4\pi G a^2) \delta p,
\end{eqnarray}
where the prime denotes conformal time derivatives, $\mathcal{H}=aH$ and $c^2_a = p' /\rho'$ denotes the adiabatic speed\footnote{This definition is also preserved in physical time units.}, where $p$ and $\rho$ are the pressure and density, respectively. $\delta p$ are the perturbations related to the pressure of the matter field. If we consider Eq.(\ref{eq:bardeen}) in the standard perturbed Einstein field equations at first order we can recover
\begin{eqnarray}
\label{eq:bardeen2}
&& \Psi'' -3\mathcal{H} (1+c^2_a) \Psi' -c^2_a \nabla^2\Psi 
+ [2\mathcal{H}' + (1+3c^2_a)\mathcal{H}^2]\Psi 
= \left( c^2_s -c^2_a\right) \nabla^2 \Psi,
\end{eqnarray}
where $c_s$ is referred to the speed of the perturbations. If we compare Eq.(\ref{eq:bardeen}) and Eq.(\ref{eq:bardeen2}) we can derive the expression that relates both speeds of the perturbations as:
\begin{eqnarray}
\label{eq:pertb-sound}
\delta p = \left(\frac{ c^2_s -c^2_a}{4\pi G a^2}\right)  \nabla^2 \Psi.
\end{eqnarray}
By using the conservation equation $\rho' + 3\mathcal{H} (\rho + p)=0$, derived from the energy momentum tensor $T^{\mu}_{\nu} =(\partial \mathcal{L}/\partial X) (\partial^\mu \phi \partial_\nu \phi) -\delta^{\mu}_{\nu} \mathcal{L}$, with $X=\frac{1}{2} \partial_\mu \phi \partial^\mu \phi$, we can obtain
\begin{eqnarray}
c^2_a = \left(\frac{p'}{\rho'} \right) &=& \left( \frac{\dot{p}}{3H (\rho + p)} \right)
=-\left( \frac{\mathcal{L}_\phi+\ddot{\phi} \mathcal{L}_X}{3H\dot{\phi} \mathcal{L}_X} \right).
\end{eqnarray}
We performed a change rule to transform to physical time, and the subindex denotes derivatives concerning $X$ and $\phi$. And 
\begin{eqnarray}
c^2_s = \frac{\left(\frac{\partial\mathcal{L}}{\partial X}\right)}{\frac{\partial\mathcal{L}}{\partial X}+ 2X \left(\frac{\partial^2\mathcal{L}}{\partial X^2}\right)}.
\end{eqnarray}

To have a perfect fluid described by an equation-of-state in the background, we need to have a vanishing $\delta p$ in the perturbative scheme, therefore, to achieve such a scenario we notice from Eq.(\ref{eq:pertb-sound}) that $c^2_s = c^2_a$, this is the so-called non-adiabatic condition. 

The latter condition implies that \cite{Unnikrishnan:2010ag}
\begin{eqnarray}
\label{eq:pertb-cond}
\frac{\partial}{\partial X} \left[ \frac{\mathcal{L}_\phi}{X \mathcal{L}_X}  \right] =0,
\end{eqnarray}
therefore, to satisfy the condition for a perfect fluid form like we need to consider, at the perturbative level, the following geometrical perturbed test: \\

4. \textit{If Eq.(\ref{eq:pertb-cond}) is satisfied, therefore $\phi$ is a perfect scalar field.}


\section{Geometric perfect fluid from $f(R)$}
\label{sec:geomPF}

Our next step is to follow the analogies from the statements \textit{1-4}, in the context of $f(R)$ gravity. First, we need to verify if the high derivatives of $R$ in this scheme are fulfilled according to \textit{1-3}, this will denote that the extra terms rising from the geometrical part can be associated with a perfect fluid form by comparing them with the standard matter terms. Second, at the perturbative level, we need to derive the equations related to the non-adiabatic condition in $f(R)$ and compute the possible constrictions on a $f(R)$ smooth function to satisfy a modified version of Eq.(\ref{eq:pertb-cond}), and therefore fulfill the condition \textit{4}.

\subsection{Background conditions: perfect fluid in $f(R)$ gravity}

We need to consider an effective fluid approach to add fluids in extended theories from Einstein's gravity, as $f(R)$ landscape. Many works have been done in this direction \cite{Tsujikawa:2007gd,Escamilla-Rivera:2016aca,Bahamonde:2017ize,Nesseris:2022hhc}, are references cited in there. However, it is standard in all of them, to begin with, a fluid of the form $\rho = \rho_{\text{fld}} + \rho{_m}$, where $\rho_{\text{fld}}$ is usually associated with dark energy and $m$ already includes the dark matter rate in the baryonic component. In this context, we specify the modified Einstein-Hilbert action as:
\begin{eqnarray}
\label{eq:actionfR}
S=\int d^{4}x\sqrt{-g}\left[  \frac{1}{2\kappa}f\left(  R\right)
+\mathcal{L}_\mathrm{m}\right],  
\end{eqnarray}
where $\mathcal{L}_\mathrm{m}$ is the matter contribution Lagrangian and
$\kappa\equiv8\pi G$, where $G$ is the Newton's constant. Varying Eq.~\eqref{eq:actionfR} with respect to the metric $g_{\mu\nu}$, we can derive the following field equations:
\begin{eqnarray}
\label{eq:fieldfR}
f_R G_{\mu\nu}-\frac{1}{2} \left[f-R f_R\right] g_{\mu\nu}+
\left(g_{\mu\nu} \Box-\nabla_\mu\nabla_\nu\right)f_R =\kappa\,T_{\mu\nu}^{(m)}, \quad 
\end{eqnarray}
where $G_{\mu\nu}$ is the standard Einstein tensor, $f_R =\partial f/\partial R$, and $T_{\mu\nu}^{(m)}$ is the energy-momentum tensor of the matter fields, which is one contribution of a general $T_{\mu\nu}= T_{\mu\nu}^{(m)} +T_{\mu\nu}^{(\text{fld})}$. According to this latter, we can consider that $T_{\mu\nu}^{(\text{fld})}$ can be a contribution related to the geometrical part of Eq.(\ref{eq:fieldfR}) as
\begin{eqnarray}
\label{eq:Tfld}
\kappa T_{\mu\nu}^{(\text{fld})}\equiv [1-f_R]G_{\mu\nu}+\frac12[f-R f_R] g_{\mu\nu} 
-\left(g_{\mu\nu}\Box-\nabla_\mu\nabla_\nu\right)f_R, \quad
\end{eqnarray}
which satisfy the conservation equation $\nabla^\mu T_{\mu\nu}^{(\text{fld})}=0$.
The evolution equations associated to Eq.(\ref{eq:fieldfR}) considering Eq.(\ref{eq:Tfld}) are given by
\begin{eqnarray}
3\mathcal{H}^2&=&\kappa a^2 \left(\rho_\mathrm{m}+ \rho_\mathrm{fld}\right), \\
6 \dot{\mathcal{H}}&=&-\kappa a^2 \big[\left(\rho_\mathrm{m}+3 p_\mathrm{m}\right)+\left(\rho_\mathrm{fld}+3 p_\mathrm{fld}\right)\big],
\end{eqnarray}
where
\begin{eqnarray}
\kappa \rho_\mathrm{fld}&=&-\frac{f}2+3\frac{\mathcal{H}^2}{a^2} -3\frac{\mathcal{H}\dot{f_R}}{a^2} +3\frac{f_R \dot{\mathcal{H}}}{a^2},
\label{eq:denfld}\\
\kappa p_\mathrm{fld}&=&\frac{f}2- \frac{\mathcal{H}^2}{a^2} -\frac{2f_R \mathcal{H}^2}{a^2} + \frac{\mathcal{H}\dot{f_R}}{a^2} -\frac{\dot{\mathcal{H}}}{a^2}-\frac{f_R\dot{\mathcal{H}}}{a^2}+\frac{\ddot{f_R}}{a^2}, \quad\quad \label{eq:pfld}
\end{eqnarray}
Using Eqs.~(\ref{eq:denfld}) and (\ref{eq:pfld}) we can derive the equation-of-state (EoS) for the fluid as:
\begin{eqnarray}
\label{eq:wfld}
&& w_\mathrm{fld} =\frac{ p_\mathrm{fld}}{\rho_\mathrm{fld}} 
\nonumber\\ &&
=\frac{2\left[(1+2f_R)\mathcal{H}^2-\mathcal{H}\dot{f_R}+(2+f_R)\dot{\mathcal{H}}-\ddot{f_R}\right] -a^2 f}{a^2 f-6(\mathcal{H}^2-\mathcal{H}\dot{f_R}+f_R\dot{\mathcal{H}})}, \quad\quad
\end{eqnarray}
which for $f(R)=R$, we obtain the standard EoS from Einstein's gravity \cite{Copeland:2006wr}. Furthermore, if we consider $\dot{R}=0$ and a flat space $k=0$, we obtain the Eos for a perfect fluid $p=w \rho $.
Following this idea and the conditions from \textit{3 -1}, Eq.(\ref{eq:fieldfR}) describe perfect fluids through the contributions from the geometric part in $f(R)$.

\subsection{Scalar perturbation conditions: non-adiabatic pressure in $f(R)$}
\label{subsec:perturbations2}

Since the contribution due to $T_{\mu\nu}^{(m)}$ is effectively associated with a standard perfect fluid, in such case with the standard matter, and we proved that $T_{\mu\nu}^{(\text{fld})}$ can describe a perfect fluid concerning the extra geometric contributions, then we are ready to demonstrate is this condition is fulfilled at the perturbative level. 
Full analyses on $f(R)$ scalar perturbations have been presented in \cite{Clifton:2011jh,Duniya:2022vdi}. In this work, we redo these calculations to obtain the quantities to satisfy the condition \textit{4}.

We start with the gravitational field constraint equations given by 
\begin{eqnarray}
&&\Psi' + {\cal H}\Phi = -\dfrac{1}{2}\kappa^2 a^2\sum \left(\rho + p\right) V, \label{eq:pert-eqs1}\\
&& \nabla^{2} \Psi - 3\mathcal{H} \left( {\cal H}\Phi +{\Psi}'\right) = \dfrac{1}{2}\kappa^2 a^2 \sum{ \delta\rho} \label{eq:pert-eqs2},
\end{eqnarray}
where $\rho$ and $p$ denote the density and pressure of all the fluid/matter components, in our case, matter (m) and the fluid (fld), and $V$ is the effective velocity potential. Since we are associated our perfect fluid with the geometrical contribution from a $f(R)$ gravity, we define in conformal time
\begin{eqnarray}
\label{eq:Vfld}
\kappa^2(\rho_\mathrm{fld}+p_\mathrm{fld}) V_\mathrm{fld} &{\equiv}& a^{-2}\left(\Phi f'_R + {\cal H}\delta{f}_R - \delta{f}'_R\right) 
+\; 2a^{-2}\left(\Psi' + {\cal H}\Phi\right)f_R , 
\end{eqnarray}
where $\rho_\mathrm{fld}$ and $p_\mathrm{fld}$ are given by Eqs.(\ref{eq:denfld})-(\ref{eq:pfld}). The fluid density perturbation can be derived as
\begin{eqnarray}
\label{eq:Vfld2}
\kappa^2\delta\rho_\mathrm{fld} &{\equiv}& 2a^{-2}\left[3{\cal H}(\Psi'+{\cal H}\Phi) - \nabla^2\Psi\right]f_R 
+\; a^{-2}\left(\nabla^2 + 3{\cal H}'\right)\delta{f}_R - 3a^{-2}{\cal H}\delta{f}'_R 
+\; 3a^{-2}\left(\Psi'+2{\cal H}\Phi\right)f'_R.
\end{eqnarray}
In order to arrive to an analogous expression as Eq.\eqref{eq:pertb-sound}, we can combine Eq.\eqref{eq:pert-eqs1}-\eqref{eq:pert-eqs2} to obtain
\begin{eqnarray}
\nabla^2\Psi \;=\; \dfrac{1}{2}\kappa^2 a^2 \sum \rho \xi ,
\end{eqnarray}
where $\xi = \Omega_m\xi_m + \Omega_\mathrm{fld}\xi_\mathrm{fld}$, denotes the comoving overdensity with 
\begin{eqnarray}
\kappa^2 a^2 \rho_\mathrm{fld} \xi_\mathrm{fld} &=& 3\left(\Psi'+{\cal H}\Phi\right)f'_R - 2f_R\nabla^2\Psi 
+\; \left[\nabla^2+3\left({\cal H}'-{\cal H}^2\right)\right]\delta{f}_R,
\end{eqnarray}
by using Eqs.\eqref{eq:Vfld}-\eqref{eq:Vfld2}. As in \eqref{eq:bardeen}, we can establish the relation between Bardeen potentials through $\Psi -\Phi = \kappa^2 a^2 \sum \left(\rho + p\right)\Pi$, where $\Pi$ denotes the anisotropic stress or anisotropic pressure dimensionless and define the evolution equation for the fluid as
\begin{eqnarray}
 \label{Pixdensity}
\kappa^2 \left[\delta{p}_\mathrm{fld} + \dfrac{2}{3}\left(\rho_\mathrm{fld}+p_\mathrm{fld}\right) \nabla^2\Pi_\mathrm{fld}\right]  &{\equiv}& \; a^{-2}\Big\{ \delta{f}''_R + {\cal H}\delta{f}'_R - 2\Phi f''_R
- \left(4\nabla^2 +{\cal H}^2 + a^2R\right)\dfrac{\delta{f}_R}{6}  
- 2f_R\Big[\Psi'' + 2{\cal H} \Psi' + {\cal H}\Phi'  \nonumber\\
&& +\; \left({\cal H}^2 + 2{\cal H}'\right) \Phi
+\; \dfrac{1}{3}\nabla^2 (\Phi-\Psi)\Big] 
- \left[\Phi' + 2(\Psi' + {\cal H}\Phi)\right]f_R' \Big\},
\end{eqnarray}
and using \eqref{eq:bardeen} we obtain
\begin{eqnarray}
\Psi'' + {\cal H} \left(2 + 3c^2_{a}\right)\Psi' +&\; {\cal H}\Phi' + \left[2{\cal H}' + \left(1+3c^2_{a}\right){\cal H}^2\right] \Phi 
=\; \dfrac{1}{3}\nabla^2\left(\Psi - \Phi\right) + \dfrac{3}{2} {\cal H}^2 c^2_{s} \xi,
\end{eqnarray}
where
\begin{eqnarray}
\label{delPx}
\delta{p} = c^2_{a}\delta\rho + \left(c^2_{s}-c^2_{a}\right) \rho \xi.
\end{eqnarray}
Under the non-adiabatic condition we recover from the latter that $c^2_s = c^2_a$, henceforth, by assuming pressureless matter ($p_m = \delta{p}_m =0$), the fluid \textit{fld} comes from a geometrical $f(R)$ contribution. Furthermore, such scenarios have been directly related to barotropic perfect fluids, i.e with vanishing non-adiabatic pressure perturbations. Interesting scenarios on this matter have been discussed in \cite{Unnikrishnan:2010ag}. However, this assumption is restricted to standard gravity models. In this work, we extended this assumption by coupling minimally a scalar field to the action \eqref{eq:actionfR}. We will adopt this analysis from a chameleon scalar field scheme in what follows.


\section{Chameleon perfect scalar field in $f(R)$}
\label{sec:chameleon}

In the context of scalar field-driven expansion dynamics, e.g. inflation epoch, a possible candidate so-called chameleon scalar field \cite{Khoury:2003aq,Khoury:2003rn,Waterhouse:2006wv,Khoury:2013yya,Saba:2017xur} has been suggested as to drive an inflationary expansion.
One particular characteristic of this scalar field is that its mass depends on the matter density effects, i.e a scalar field with a varying mass in a dense scenario, where the scalar field can acquire a large mass in a short range. Some observational tests have been performed at CMB scales \cite{Sheikhahmadi:2019gzs} to associate this scalar field to an early accelerated expansion phase.
In the context of $f(R)$, the chameleon scalar field has been studied in \cite{Burrage:2017qrf,Brax:2008hh}. However, if the chameleon scalar field can have the property a perfect scalar needs to be proved. 

To develop the latter, we are going to consider the following action:
\begin{eqnarray}
\label{eq:actionfR-phi}
S=\int d^{4}x\sqrt{-g} \frac{1}{2} \left[  M^2_{\text{pl}} f(R) - \partial_{\mu} \phi \partial^{\mu} \phi - 2V(\phi) 
+\mathcal{L}_\mathrm{m}\right].  \quad 
\end{eqnarray}
In this landscape, we are going to consider a conformal transformation to relate this action with a standard one in a scalar-tensor theory through
\begin{equation}
\label{eq:transformation}
\exp\left(-\frac{2\beta \phi}{M_{\text{pl}}}\right)=f^{\prime}(R).
\end{equation}
where $\beta=\sqrt{\frac{1}{6}}$ \cite{Brax:2008hh}. In Einstein frame we can write the metric as $\bar{g}_{\mu\nu}$ by a conformal transformation defined as:
\begin{equation}\label{g}
\bar{g}_{\mu\nu}=e^{-\frac{2\beta \phi}{M_{\text{pl}}}}g_{\mu\nu},
\end{equation}
therefore, we can rewrite (\ref{eq:actionfR-phi}) as
\begin{eqnarray}
S=\int d^4x\sqrt{-g} \left(\frac{R}{2\kappa} -\frac{1}{2} \partial_{\mu}\phi \partial_{\nu} \phi-V(\phi)\right)
+S_{\mathrm{m}}[e^\frac{2\beta \phi}{M_{\text{pl}}}], \quad 
\end{eqnarray}
where 
\begin{equation}
\label{eq:v}
V(\phi)=\frac{M_{\text{pl}}^2 \left[Rf^{\prime}(R)-f(R)\right]}{2f^{\prime}(R)^2}.
\end{equation}
Notice that $f(R)$ theories are equivalent to scalar-tensor theories in this chameleon scheme. Now, from the conditions described in Sec.\ref{sec:geomPF}, we can compute the explicit form for the chameleon geometric scheme in $f(R)$.
First, we need to consider a Lagrangian density of the form
\begin{eqnarray}
\label{eq:L}
\mathcal{L}(X,\phi) = f(X) -V(\phi).
\end{eqnarray}
While a form of $f(X)\propto \log(X)$ satisfies directly the condition \textit{4}, in this scheme we need to verify this condition on the chameleon potential.
In a $f(R)$ theory to have a chameleon mechanism it is required that the derivatives of the potential behave as:
$V^{\prime}(\phi) < 0, ~ V^{\prime \prime}(\phi)> 0, ~\text{and,} ~ V^{\prime \prime \prime}(\phi) < 0.$
From (\ref{eq:v}) we can derive
\begin{eqnarray}
V^{\prime}(\phi) &=& \frac{\beta M_\text{pl}}{f^{\prime\, 2}}\left[R f^{\prime}- 2f\right], \label{eq:der-poten1} \\
V^{\prime \prime}(\phi)&=& \frac{1}{3}\left[\frac{R}{f^{\prime}}+\frac{1}{f^{\prime \prime}} - \frac{4f}{f^{\prime\,2}}\right], \label{eq:der-poten2}\\
V^{\prime \prime \prime}(\phi) &=& \frac{2\beta}{3 M_\text{pl}}
\left[\frac{3}{f^{\prime \prime}} + \frac{f^{\prime}f^{\prime
\prime \prime}}{f^{\prime \prime\,3}}+\frac{R}{f^{\prime}} -
\frac{8f}{f^{\prime\,2}}\right]. \label{eq:der-poten3}
\end{eqnarray}
Generally, these functions give tight constraints on the form selected for $f(R)$, therefore for a specific form of $V(\phi)$ we can find a specific $f(R)$.
To perform this calculation, a potential that follows the chameleon mechanism given by Eqs.(\ref{eq:der-poten1})-(\ref{eq:der-poten2})-(\ref{eq:der-poten3}) is a power law potential of the form:
\begin{eqnarray}
V(\phi) = M^{4+n} \phi^{-n},
\end{eqnarray}
and introduce this expression in Eq.(\ref{eq:v}) and solve the differential equation to obtain an exact form for $f(R)$:
\begin{eqnarray}
\label{eq:f-poten}
f(R)=\frac{A+2^{2/3} B-32 M^{n+4} M_{\text{pl}}^4-32 M_{\text{pl}}^4 \phi ^n+4 R^2 M_{\text{pl}}^8}{16 M_{\text{pl}}^6}, \quad
 \end{eqnarray}
where
   \begin{eqnarray}
A&=& \frac{2 \sqrt[3]{2} R M_{\text{pl}}^{10} \left(R^3 M_{\text{pl}}^6+2\right)}{B}, \\
B&=& {(C-2 R^6 M_{\text{pl}}^{24}+10 R^3 M_{\text{pl}}^{18}+M_{\text{pl}}^{12})}^{1/3}, \\
C&=& \sqrt{-M_{\text{pl}}^{24} \left(4 R^3 M_{\text{pl}}^6-1\right){}^3}.
 \end{eqnarray}
 Eq.(\ref{eq:f-poten}) is the specific $f(R)$ form for a perfect chameleon scalar field. Also, we can perform the derivation of this latter expression to obtain:
 \begin{eqnarray}
 \label{eq:fRder}
f_{R}= \frac{1}{8} M_{\text{pl}}^2 (f_1 + f_2- f_3 + f_4 +4 R),
 \end{eqnarray}
 where
 \begin{eqnarray}
 f_1 &=& \frac{3 \sqrt[3]{2} R^3 M_{\text{pl}}^8}{\sqrt[3]{-2 R^6 M_{\text{pl}}^{24}+10 R^3 M_{\text{pl}}^{18}+\sqrt{-M_{\text{pl}}^{24} \left(4 R^3
   M_{\text{pl}}^6-1\right){}^3}+M_{\text{pl}}^{12}}}, \\
   f_2 &=& \frac{\sqrt[3]{2} M_{\text{pl}}^2 \left(R^3 M_{\text{pl}}^6+2\right)}{\sqrt[3]{-2 R^6 M_{\text{pl}}^{24}+10 R^3 M_{\text{pl}}^{18}+\sqrt{-M_{\text{pl}}^{24} \left(4
   R^3 M_{\text{pl}}^6-1\right){}^3}+M_{\text{pl}}^{12}}} ,\\
   f_3 &=& \frac{2 \sqrt[3]{2} R^3 M_{\text{pl}}^{20} \left(R^3 M_{\text{pl}}^6+2\right) \left(-2 R^3 M_{\text{pl}}^6-\frac{3 \sqrt{-M_{\text{pl}}^{24} \left(4 R^3
   M_{\text{pl}}^6-1\right){}^3}}{M_{\text{pl}}^{12}-4 R^3 M_{\text{pl}}^{18}}+5\right)}{\left(-2 R^6 M_{\text{pl}}^{24}+10 R^3
   M_{\text{pl}}^{18}+\sqrt{-M_{\text{pl}}^{24} \left(4 R^3 M_{\text{pl}}^6-1\right){}^3}+M_{\text{pl}}^{12}\right){}^{4/3}}, \\
   f_4&=& \frac{2^{2/3} R^2 M_{\text{pl}}^{10} \left(-2 R^3 M_{\text{pl}}^6-\frac{3 \sqrt{-M_{\text{pl}}^{24} \left(4 R^3 M_{\text{pl}}^6-1\right){}^3}}{M_{\text{pl}}^{12}-4
   R^3 M_{\text{pl}}^{18}}+5\right)}{\left(-2 R^6 M_{\text{pl}}^{24}+10 R^3 M_{\text{pl}}^{18}+\sqrt{-M_{\text{pl}}^{24} \left(4 R^3
   M_{\text{pl}}^6-1\right){}^3}+M_{\text{pl}}^{12}\right){}^{2/3}}.
 \end{eqnarray}
 Introducing Eqs.(\ref{eq:f-poten})-(\ref{eq:fRder}) in Eq.(\ref{eq:Tfld}), we obtain for the time-time component:
 \begin{eqnarray}
 \label{eq:0-0}
 -6 \kappa  \left(g_3+g_4-g_5+g_6+4 R\right) M_{\text{pl}}^2 \left(3 p_{\text{fld}}+\rho _{\text{fld}}\right)+32 \kappa  \rho _{\text{fld}}  = \frac{3
   \left(g_1+2^{2/3} g_2-32 M^{n+4} M_{\text{pl}}^4-32 M_{\text{pl}}^4 \phi ^n+4 R^2 M_{\text{pl}}^8\right)}{M_{\text{pl}}^6},
 \end{eqnarray}
 where
 \begin{eqnarray}
 g_1 &=& \frac{2 \sqrt[3]{2} R M_{\text{pl}}^{10} \left(R^3 M_{\text{pl}}^6+2\right)}{\sqrt[3]{-2 R^6 M_{\text{pl}}^{24}+10 R^3 M_{\text{pl}}^{18}+\sqrt{-M_{\text{pl}}^{24} \left(4 R^3
   M_{\text{pl}}^6-1\right){}^3}+M_{\text{pl}}^{12}}}, \\
   g_2 &=& \sqrt[3]{-2 R^6 M_{\text{pl}}^{24}+10 R^3 M_{\text{pl}}^{18}+\sqrt{-M_{\text{pl}}^{24} \left(4 R^3 M_{\text{pl}}^6-1\right){}^3}+M_{\text{pl}}^{12}}, \\
   g_3&=& \frac{3 \sqrt[3]{2} R^3 M_{\text{pl}}^8}{\sqrt[3]{-2 R^6 M_{\text{pl}}^{24}+10 R^3 M_{\text{pl}}^{18}+\sqrt{-M_{\text{pl}}^{24} \left(4 R^3
   M_{\text{pl}}^6-1\right){}^3}+M_{\text{pl}}^{12}}},\\
   g_4&=& \frac{\sqrt[3]{2} M_{\text{pl}}^2 \left(R^3 M_{\text{pl}}^6+2\right)}{\sqrt[3]{-2 R^6 M_{\text{pl}}^{24}+10 R^3 M_{\text{pl}}^{18}+\sqrt{-M_{\text{pl}}^{24} \left(4 R^3
   M_{\text{pl}}^6-1\right){}^3}+M_{\text{pl}}^{12}}}, \\
   g_5&=& \frac{2 \sqrt[3]{2} R^3 M_{\text{pl}}^{20} \left(R^3 M_{\text{pl}}^6+2\right) \left(-2 R^3 M_{\text{pl}}^6-\frac{3 \sqrt{-M_{\text{pl}}^{24} \left(4 R^3
   M_{\text{pl}}^6-1\right){}^3}}{M_{\text{pl}}^{12}-4 R^3 M_{\text{pl}}^{18}}+5\right)}{\left(-2 R^6 M_{\text{pl}}^{24}+10 R^3 M_{\text{pl}}^{18}+\sqrt{-M_{\text{pl}}^{24} \left(4
   R^3 M_{\text{pl}}^6-1\right){}^3}+M_{\text{pl}}^{12}\right){}^{4/3}}, \\
   g_6&=& \frac{2^{2/3} R^2 M_{\text{pl}}^{10} \left(-2 R^3 M_{\text{pl}}^6-\frac{3 \sqrt{-M_{\text{pl}}^{24} \left(4 R^3 M_{\text{pl}}^6-1\right){}^3}}{M_{\text{pl}}^{12}-4 R^3
   M_{\text{pl}}^{18}}+5\right)}{\left(-2 R^6 M_{\text{pl}}^{24}+10 R^3 M_{\text{pl}}^{18}+\sqrt{-M_{\text{pl}}^{24} \left(4 R^3
   M_{\text{pl}}^6-1\right){}^3}+M_{\text{pl}}^{12}\right){}^{2/3}}.
 \end{eqnarray}
Notice that the expression (\ref{eq:0-0}) and its second order derivative satisfy the conditions from \textit{1 -3}, therefore the chameleon field $\phi$ with potential Eq.(\ref{eq:v}) describe a perfect fluid through the contributions from the geometric part in $f(R)$.

 \section{Discussion}
 \label{sec:discussions}
 
In this work, we considered a particular class of chameleon field $\phi$ with a power-law potential of the form $V(\phi) = M^{4+n} \phi^{-n}$. Under this assumption, we found that the chameleon can be associated with a perfect scalar field since its behaviour on the evolution and conservative background equations satisfy the conditions described by \textit{1 -3}. Furthermore, at the perturbative level, $\phi$ fulfill directly the condition \textit{4} if we consider a Lagrangian of the form (\ref{eq:L}). For $n\geq 1$, we notice from Eq.(\ref{eq:0-0}) that local gravity constraints are satisfied for M $\sim 10^{-2}$eV. As we mentioned, this is a restricted condition 
on the energy scale required for the current cosmic acceleration observed. While the conditions at the background evolution seem a \textit{natural} candidate for dark energy for this kind of potential, it is interesting to notice that at a perturbative level, a chameleon perfect scalar $\phi$ could emerge at first during a phase transition during the inflation era and, finally end with a random position within its potential. Following this evolution, there has been a study on attractors conditions in which the chameleon $\phi$ could satisfy the big bang nucleosynthesis (BBN) constraints until today \cite{Mota:2011nh}, which behind the idea of a perfect scalar field at perturbative level can help to set up an ensemble of initial conditions to study furthermore its behaviour at the early universe. This study will be reported elsewhere.

 
\acknowledgments
LLP is supported by CONACyT National Grant.
CE-R acknowledges the Royal Astronomical Society as FRAS 10147 and is supported by DGAPA-PAPIIT-UNAM Project TA100122.
This article is based upon work from COST Action CA21136 Addressing observational tensions in cosmology with systematics and fundamental physics (CosmoVerse) supported by COST (European Cooperation in Science and Technology).


\bibliographystyle{utphys}
\bibliography{references}


\end{document}